\documentclass{moriond}

\def\Journal#1#2#3#4{{#1} {\bf #2}, #3 (#4)}

\def\PRL{\em Phys. Rev. Lett.}

\def\JHEP{{\em JHEP}}

\def\be{\begin{equation}}
\def\ee{\end{equation}}
\def\bea{\begin{eqnarray}}
\def\eea{\end{eqnarray}}

\newcommand{\Photo}{}

\usepackage{amsmath}
\usepackage{placeins}

\begin{document}
\vspace*{-1cm}
\title{Beautiful mixing and CP violation at LHCb}

\author{ Philippe d'Argent on behalf of the LHCb collaboration }

\address{CERN, Geneva, Switzerland}

\maketitle\abstracts{
Precision measurements of beauty hadron decays are sensitive probes of the Standard
Model and a promising way to look for new physics phenomena
far beyond the energy scale accessible for
direct production searches.
This article reviews recent measurements of mixing and CP violation in beauty decays 
performed at the LHCb experiment
that have been presented at 
the $55^{th}$ Rencontres de Moriond QCD conference.
}

\vspace{-0.5\baselineskip}
\section{The Standard Model and beyond}
\vspace{-0.5\baselineskip}

\sloppypar
In the framework of the Standard Model of particle physics, the charge-parity (CP) symmetry between quarks and antiquarks is broken by a single complex
phase in the Cabibbo-Kobayashi-Maskawa (CKM) quark-mixing matrix. 
The unitarity of this matrix leads to the condition $V_{ud}^{\phantom{*}}V^{*}_{ub} + V_{cd}^{\phantom{*}}V^{*}_{cb} + V_{td}^{\phantom{*}}V^{*}_{tb} = 0$, where $V_{ij}$ are the complex elements of the CKM matrix.
This equation can be visualised as a triangle in the complex plane
with angles $\alpha$, $\beta$ and~$\gamma$.
A key consistency test of the Standard Model is to verify the unitarity conditions
by over-constraining the CKM matrix with various independent measurements sensitive to
distinct combinations of matrix elements.
While the magnitudes of the CKM matrix elements can be determined from the decay
rates of respective flavour-changing transitions, measurements of CP asymmetries generally permit
determining the CKM phases.
Here, the angle 
$\gamma \equiv \rm{arg}[-(V_{ud}^{\phantom{*}}V_{ub}^{*})/(V_{cd}^{\phantom{*}}V_{cb}^{*})]$
has particularly interesting features as it is the only
CKM angle that can be measured in tree-level decays.
In such decays, the interpretation of physical observables (rates and CP asymmetries) in terms
of the underlying CKM parameters is subject to negligible theoretical uncertainties.
Hence, a precision measurement of $\gamma$ provides a Standard
Model benchmark, to be compared with indirect determinations from 
other CKM matrix observables which are more susceptible to new physics phenomena
beyond the Standard Model.

\vspace{-0.5\baselineskip}
\section{Direct CP violation in beauty decays}
\vspace{-0.5\baselineskip}

\sloppypar
The most stringent constraints on the CKM angle $\gamma$ come from measurements
of direct CP violation in $B^\mp \to D K^\mp$ decays, where $D$ represents an admixture of the $D^0$ and $\bar D^0$
flavour states.
While the $B^- \to D^0 K^-$ decay proceeds via a $b \to c \bar u s$ quark-level transition,
a $b \to u \bar c s$ transition leads to the $B^- \to \bar D^0 K^-$ decay.
Provided that the charm meson decays into a final
state, $f$, which is accessible for both flavour states, 
phase information can be determined from the interference between these two decay paths.
The relative phase between the corresponding decay amplitudes has both CP-violating ($\gamma$)
and CP-conserving ($\delta_B^{DK}$) contributions.
A measurement of the decay rate asymmetry between $B^+$ and $B^-$ mesons thus gives
sensitivity to $\gamma$. 
The sensitivity is driven by 
the size of $r_B^{DK}$, the ratio of the magnitudes of the
$B^- \to \bar D^0 K^-$ and $B^- \to D^0 K^-$  amplitudes. 
Similar interference effects also occur in $B^\mp \to D \pi^\mp$ decays, albeit with
a significantly reduced sensitivity to the phases due to additional Cabibbo-suppression ($r_B^{DK} \approx 20 \, r_B^{D\pi}$).
Two recent measurements of direct CP violation in 
$B^\mp \to D h^\mp$ ($h \in \{K,\pi \}$) decays 
study 
two-body ($D \to h^\pm h^\mp$) and self-conjugate three-body 
($D \to K^0_{\rm s} h^\pm h^\mp$) charm decays, respectively.
Both analyses use data accumulated with
the LHCb detector over the period from 2011 to 2018 in $pp$ collisions at energies of 
$\sqrt{s} =7, 8$ and $13$ TeV, corresponding to a total integrated luminosity of approximately $9\, \rm{fb}^{-1}$.

\begin{figure}[t]
\begin{minipage}{0.6\linewidth}
\includegraphics[width=0.49\linewidth,height=4cm]{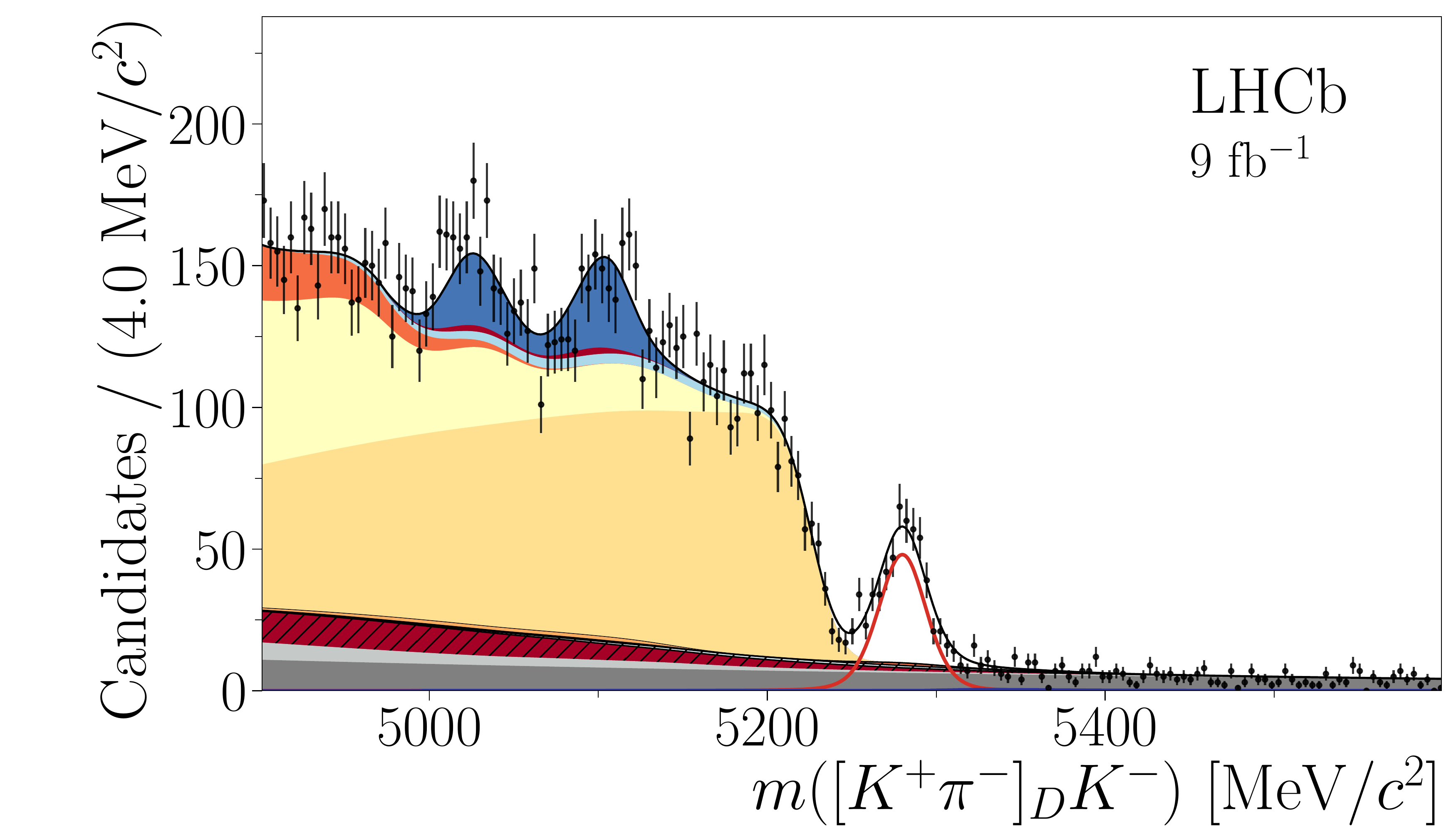}
\includegraphics[width=0.49\linewidth,height=4cm]{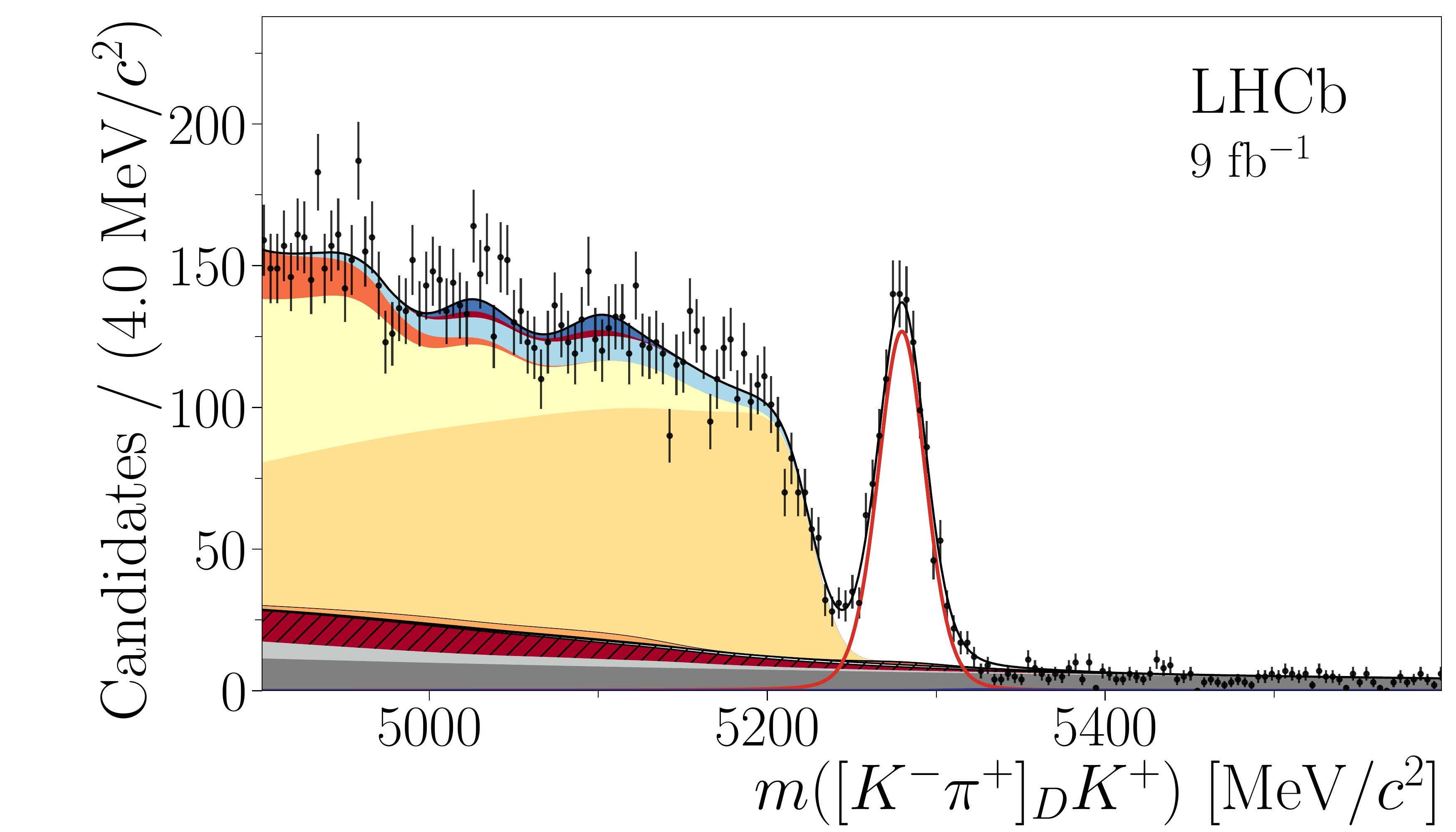}
\caption{Invariant-mass distribution of $B^- \to [K^+ \pi^-]_D K^-$ (left) and  $B^+ \to [K^- \pi^+]_D K^+$ (right) candidates with the fit projections overlaid.
The signal component (red peak) and show a huge asymmetry.  
Partially reconstructed decays are visible at low invariant mass.
}
\label{fig:1}
\end{minipage}
\hfill
\begin{minipage}{0.35\linewidth}
\includegraphics[width=\linewidth,height=4cm]{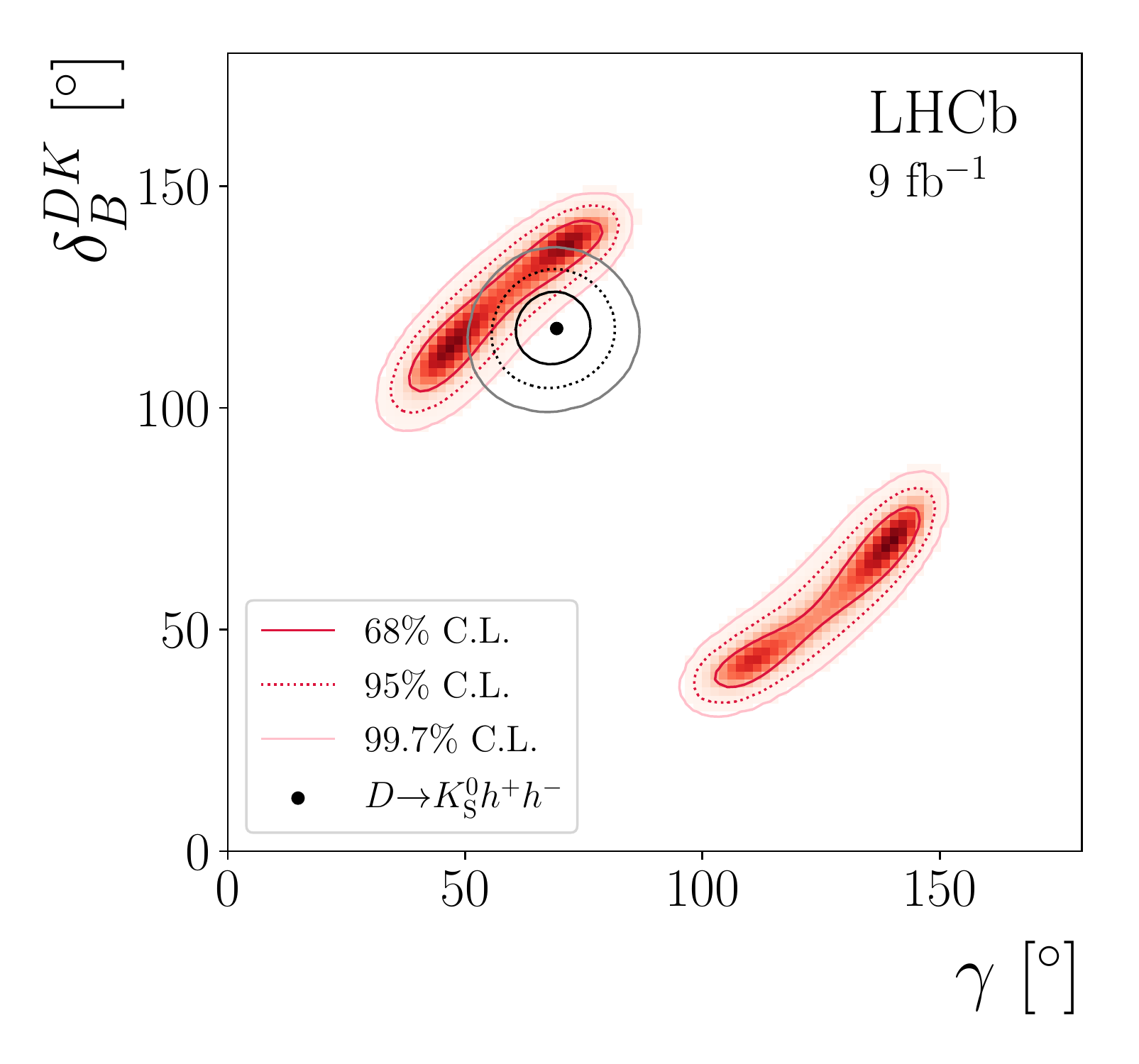}
\caption[]{
Confidence region in the ($r_B^{DK}, \gamma$) plane for both $B^\mp \to D^{(*)} h^\mp$ analyses.}
\label{fig:2}
\end{minipage}
\label{fig:radish}
\end{figure}

The first analysis~\cite{lhcb_b2dk_2body} considers the CP-eigenstates $D \to \pi^\pm \pi^\mp$
and $D \to K^\pm K^\mp$
as well as 
$D \to K^+ \pi^-$, where
the $D^0 \to K^+ \pi^-$ and $\bar D^0 \to K^+ \pi^-$
decays are related by the amplitude magnitude $r_D^{K\pi}$
and the strong-phase difference $\delta_D^{K\pi}$.
For the latter case, the
similar magnitude of $r_B^{DK}$ and $r_D^{K\pi}$
leads to significant interference between the two 
decay paths (favoured B decay followed by suppressed D decay, and suppressed
B decay followed by favoured D decay).
As is evident from the invariant-mass distributions shown in Fig.~\ref{fig:1}, this results in a huge 
asymmetry between the $B^-$ and $B^+$ decay rates. 
Moreover, the analysis includes partially reconstructed $B^\mp \to D^* h^\mp$ decays, in which the vector $D^*$ meson decays to either the $D\pi^0$ or
$D\gamma$ final state.
In total 28 observables (CP asymmetries and decay rate ratios) are measured.
The combined information allows deriving tight constrains on the underlying 
physics parameters $r_B^{X}, \delta_B^{X}, r_D^f, \delta_D^f$ and $\gamma$ ($X \in \{ DK, D\pi, D^*K, D^*\pi \}, f \in \{ K^\pm \pi^\mp, K^+K^-,\pi^+\pi^- \}$)
as displayed in Fig.~\ref{fig:2} for the ($r_B^{DK}, \gamma$) plane.

\sloppypar
Similarly, the analysis of $D \to K^0_{\rm s} h^\pm h^\mp$ decays~\cite{lhcb_b2dk_3body}
investigates differences in the phase-space distributions 
of $B^+$ and $B^-$ meson decays.
To interpret the result in terms of the physical observables, knowledge of the the strong-phase variation over the Dalitz plot of the $D$ decay is required.
A model-unbiased approach is employed that
uses direct measurements of the strong-phase difference between $D^0$ and $\bar D^0$
decays, averaged over regions of the phase space. 
These strong-phase differences have been measured by the CLEO and the BESIII
collaborations using quantum correlated pairs of $D$ mesons produced in decays of $\psi(3770)$.
The Dalitz-plot binning scheme is optimized with the help of an amplitude model. 
With this procedure, the CKM angle $\gamma$ is determined to be $\gamma = (68.7^{+5.2}_{-5.1})^\circ$, 
the most precise single measurement to date.
The results are in good agreement with the $D \to h^\pm h^\mp$ analysis
and are crucial to resolve the remaining ambiguities in the parameter space,
see Fig.~\ref{fig:2}.
\\

\sloppypar
The family of $B \to K\pi$ decays
receives significant contributions from loop-level transitions providing a powerful probe for new physics phenomena.
Measurements of direct CP violation in these channels have revealed significant deviations from the expected isospin symmetry,
an anomaly known as the $K\pi$ puzzle. 
The reconstruction of the $B^+\to K^+\pi^0$ decay is particularly challenging at a hadron collider as 
no B-meson decay vertex can be reconstructed from a single charged track.
Charged kaons that are inconsistent with originating from the primary collision point but consistent with the reconstructed trajectory of the b-meson candidate are selected
from a data sample corresponding to a luminosity of $5.4 \rm{fb}^{-1}$.
The CP asymmetry between $B^-$ and $B^+$ decay rates is found to be~\cite{lhcb_b2kpi} $A_{CP}(B^+\to K^+\pi^0) = 0.025\pm0.015\pm0.006\pm 0.003$, 
where the uncertainties are statistical, systematic and  due to external inputs.
This result is consistent with the world average and exceeds its precision.
It confirms
and significantly enhances the observed anomalous difference between the direct
CP asymmetries of the $B^+\to K^+\pi^0$ and $B^+\to K^+\pi^-$ decays. 

\vspace{-0.5\baselineskip}
\section{Mixing-induced CP violation in beauty decays}
\vspace{-0.5\baselineskip}

\begin{figure}[b]
\centering
\begin{minipage}{0.4\linewidth}
\centerline{\includegraphics[width=\linewidth]{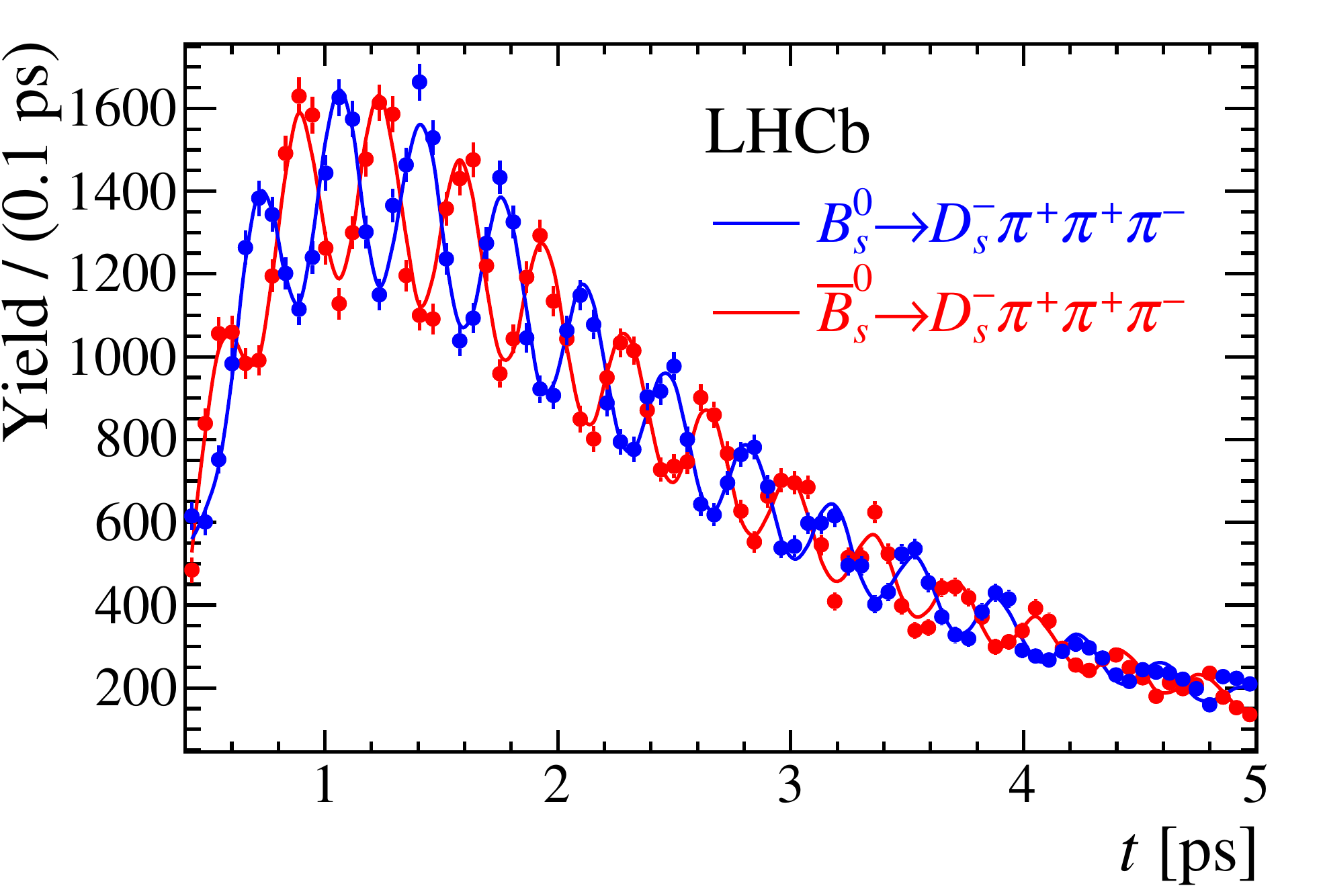}}
\end{minipage}
\hfill
\begin{minipage}{0.4\linewidth}
\centerline{\includegraphics[width=\linewidth]{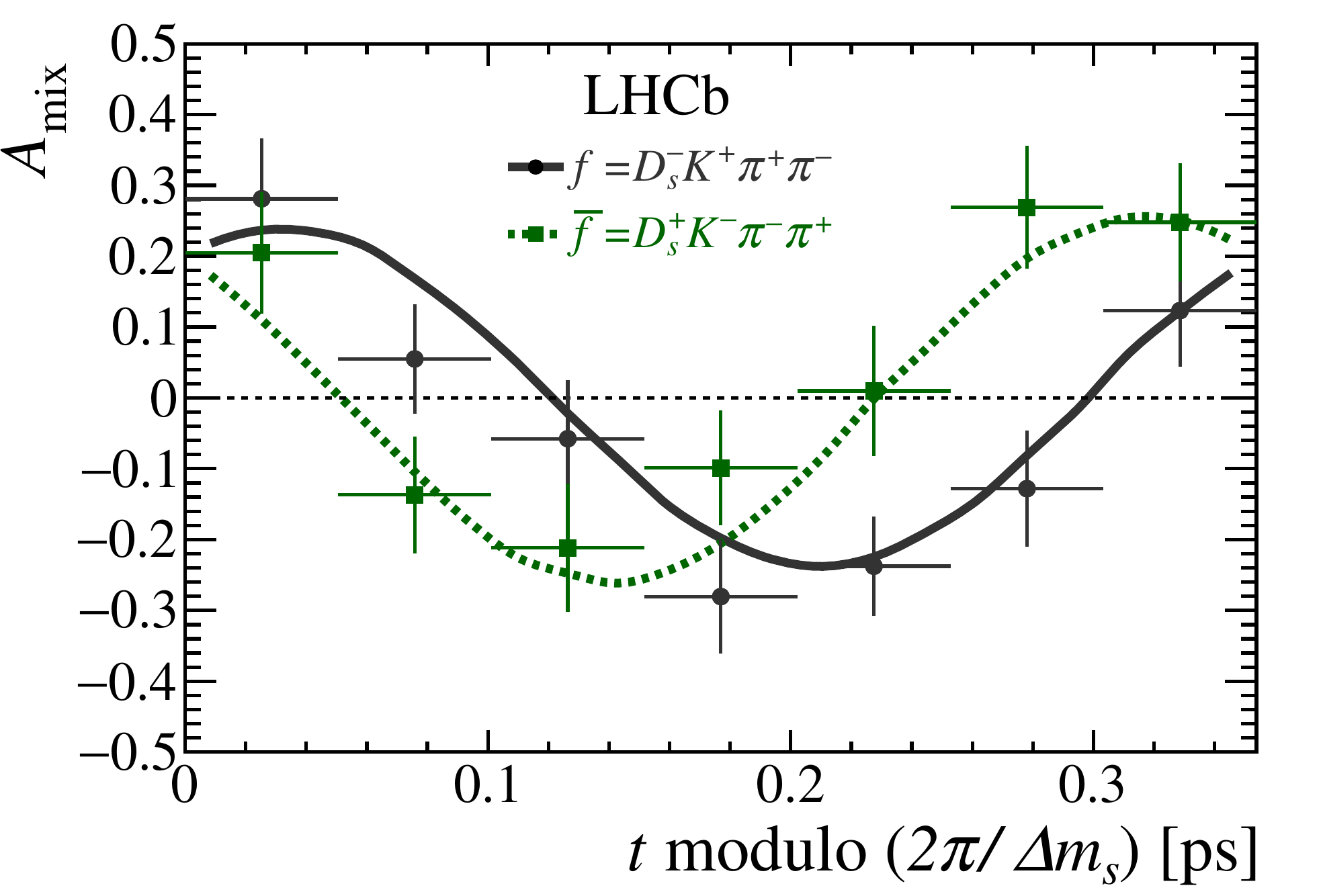}}
\end{minipage}
\caption[]{Decay-time distribution of flavour-tagged $B_s^0 \to D_s^- \pi^+ \pi^- \pi^+$ signal decays (left) and
mixing asymmetry for \mbox{$B_s^0 \to D_s^\mp K^\pm \pi^\pm \pi^\mp$} signal decays folded into one oscillation period (right).
The fit projections are overlaid (lines).
}
\label{fig:3}
\end{figure}

\sloppypar
Neutral $B_s^0$ mesons can oscillate into
their antiparticle counterparts via quantum loop processes opening additional
mechanisms for CP symmetry breaking.
The frequency of this oscillation, $\Delta m_s$, is an important parameter
of the Standard Mode and provides powerful constraints in global CKM fits.
The mixing from $B_s^0$ to $\bar B_s^0$ occurs
about three million million times per second, 
making it a major experimental challenge to resolve it.
Due to the excellent decay vertex resolution and track momentum resolution, the LHCb detector is ideally suited for this task.
Two recent measurements of $\Delta m_s$
use flavour specific $B_s^0 \to D_s^- \pi^+ \pi^- \pi^+$ ($9\, \rm{fb}^{-1}$)~\cite{lhcb_bsdskpipi}
and $B_s^0 \to D_s^- \pi^+$ ($6\, \rm{fb}^{-1}$)~\cite{lhcb_dms} decays, respectively. 
To determine if a neutral meson oscillated into its antiparticle, knowledge of the initially created flavour eigenstate is required.
This is achieved by using a combination of several flavour-tagging algorithms
that exploit different features of the b-hadron production process.
Figure~\ref{fig:3} shows the oscillation pattern 
of signal decays having the same flavour at the production and decay, and those, for which the flavour has changed. 
Both measurements of the oscillation frequency are significantly more precise than the current world-average value.
Their combination with previous LHCb measurements yields \mbox{$\Delta m_s = 17.7656 \pm 0.0057 \, \rm{ps}^{-1}$}, a crucial legacy measurement of the original LHCb detector.

\sloppypar
Interference between
the amplitudes of a $B_s^0$ meson directly decaying through $b \to  c \bar c  s$ 
into a CP eigenstate or after oscillation to a $\bar B_s^0$ meson 
gives rise to the CP violating phase $\phi_s \approx - 2 \beta_{s}$ with
$\beta_{s} \equiv \rm{arg}[-(V_{ts}^{\phantom{*}}V_{tb}^{*})/(V_{cs}^{\phantom{*}}V_{cb}^{*})]$.
The precise measurement of this phase is of high interest because of its potential sensitivity to new  
particles altering the mixing amplitudes.
A time-dependent angular analysis of $B_s^0 \to J/\psi \phi$ decays with $J/\psi \to e^+ e^-$ and $\phi \to K^+ K^-$
determines the mixing phase to be $\phi_s = 0.00 \pm 0.28 \pm 0.05 \,\rm{rad}$  ($3\, \rm{fb}^{-1}$)~\cite{lhcb_phis}.
This is the first measurement of $\phi_s$ with an electron pair in the final state. 
The result shows no evidence of CP violation and is consistent with previous measurements and the Standard
Model prediction. 
It also constitutes an important cross-check for the results with muons in the final state with independent systematic uncertainties.

\sloppypar
Complementary to the $\gamma$ measurements in charged b-hadron decays,
mixing-induced CP violation in $B_s^0 \to D_s^\mp K^\pm \pi^\pm \pi^\mp$ decays
provides sensitivity to the weak phase $\gamma - 2\beta_s$. 
This is studied for the first time
using $9\, \rm{fb}^{-1}$
of $pp$ collision data recorded by the LHCb detector~\cite{lhcb_bsdskpipi}.
Due to the multi-body final state, the hadronic parameters vary across the five dimensional phase space of the decay.
A time-dependent amplitude analysis is performed to disentangle the various intermediate-state
components contributing via $b \to c$ or $b \to u$ quark-level transitions.
The prominent contributions are found to be the 
cascade decays $B_s^0 \to D_s^\mp K_1(1270)^\pm$ and \mbox{$B_s^0 \to D_s^\mp K_1(1400)^\pm$}
with 
$K_1(1270)^\pm \to K^*(892)^0 \pi^\pm, K^\pm \rho(770)^0, K^*_0(1430)^0 \pi^\pm$ as well as 
\mbox{$K_1(1400) \to K^*(892)^0 \pi^\pm$}.
Figure~\ref{fig:3} shows the 
mixing asymmetry for final state $f=D_s^-K^+\pi^+\pi^-$, defined as
\mbox{$A_{\text{mix}}^f(t) = (N_f(t) - \bar N_f(t))/(N_f(t) + \bar N_f(t))$},
where $N_f(t)$ ($\bar N_f(t)$) denote the number of initially produced $B_s^0$ ($\bar B_s^0$) mesons.
The equivalent mixing asymmetry for the CP-conjugate process, $A_{\text{mix}}^{\bar f}(t)$, 
shows a phase shift related to the weak phase $\gamma - 2\beta_s$
signifying time-dependent CP violation.
The CKM angle $\gamma$ is determined to be $\gamma = (44\pm12)^\circ$ 
by taking the mixing phase $\beta_s$ as external input.
An alternative model-independent measurement, integrating over the phase space of the decay, gives a consistent result, $\gamma = (44^{+20}_{-13})^\circ$,
with reduced statistical precision but free of model uncertainties related to the amplitude parameterization.

\begin{figure}[t]
\vspace{-0.5\baselineskip}
\centering
\begin{minipage}{0.4\linewidth}
\includegraphics[width=\linewidth,height=4cm]{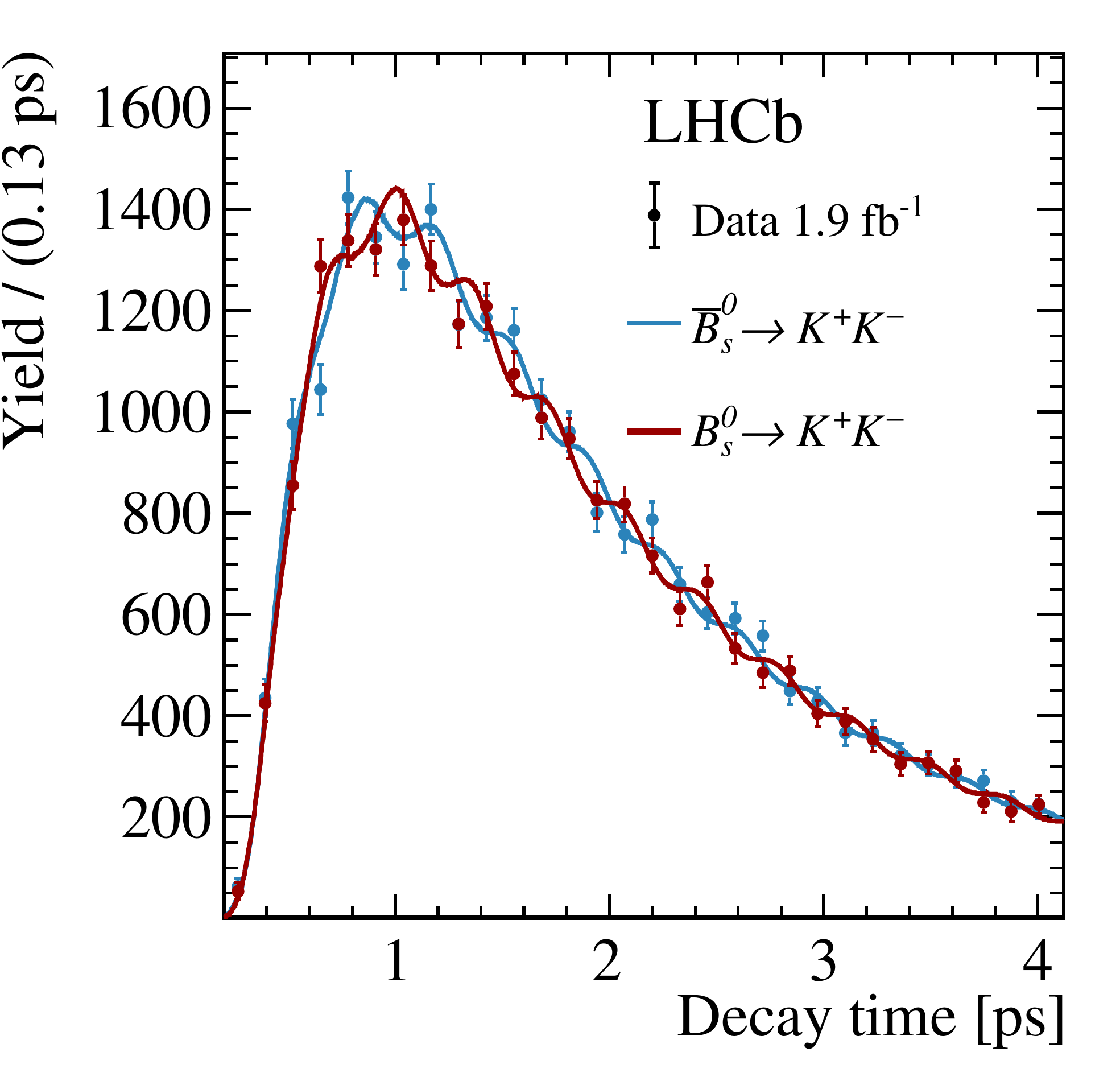}
\caption[]{
Decay-time distribution of flavour-tagged $B_s^0 \to K^+ K^-$ signal decays.
}
\label{fig:4}
\end{minipage}
\hfill
\begin{minipage}{0.4\linewidth}
\includegraphics[width=\linewidth,height=4cm]{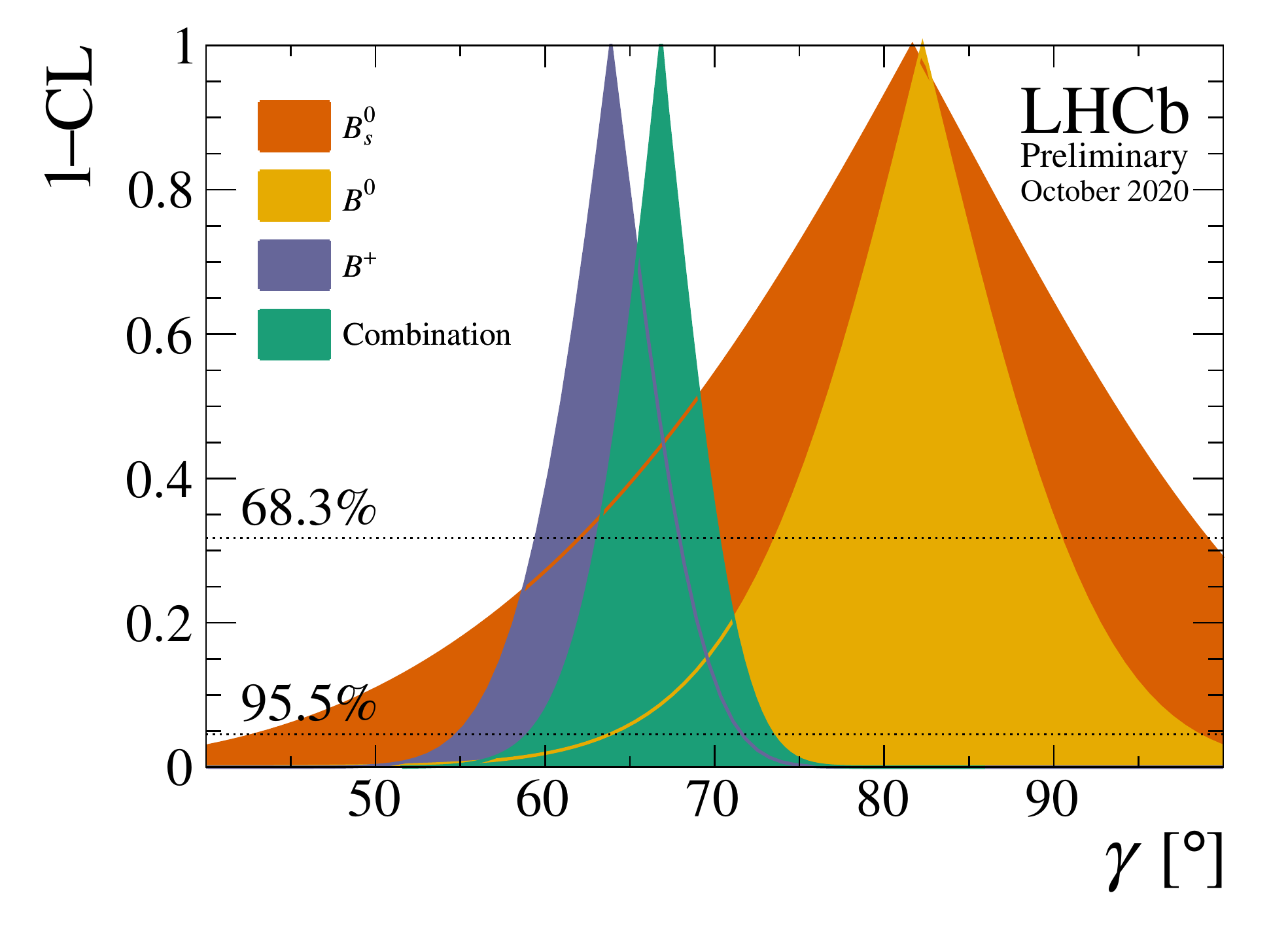}
\caption[]{
Profile-likelihood scan of 1-CL ($p$-value) for the LHCb $\gamma$ combination.
}
\label{fig:5}
\end{minipage}
\end{figure}

\sloppypar
The CP asymmetries of charmless $B^0_{(s)}$ decays to two-body charged
final states provide access to the CKM angles $\alpha$ and $\gamma$
and the $B^0$ and $B_s^0$ mixing phases.
In contrast to the tree-level measurements from $B^\mp \to D h^\mp$ and $B_s^0 \to D_s^\mp K^\pm \pi^\pm \pi^\mp$
decays, the sensitivity to the CKM angles originates from the interference of the $b\to u$ tree-level with the
$b \to d$ or $b \to s$ loop-level transitions.
Figure~\ref{fig:4} shows the decay-time distribution of flavour-tagged $B_s^0 \to K^+ K^-$ signal decays
using a data sample corresponding to a luminosity of $1.9 \rm{fb}^{-1}$.
The CP observables describing the decay-time distribution are measured with world-best precision~\cite{lhcb_b2hh}.
Combined with previous LHCb results, 
the first observation of time-dependent CP violation in the $B_s^0$ system is reported. 
This is an important milestone for flavour physics.

\vspace{-0.5\baselineskip}
\section{Outlook}
\vspace{-0.5\baselineskip}

\sloppypar
The LHCb collaboration continues to push the frontier of heavy flavour physics.
New measurements of the $B_s^0-\bar B_s^0$ mixing frequency have reached unprecedented precision. 
While time-dependent CP violation in the $B_s^0$ system has now been observed for the first time,
the breaking of CP symmetry has still not been observed in the baryon sector.
With the first amplitude analysis of any b-baryon decay mode allowing for CP-violation effects, 
the LHCb collaboration is also pioneering in this field.
Within the current precision, no significant CP asymmetries have been observed for the amplitude components
contributing to $\Xi_b^- \to p K^- K^-$ decays~\cite{lhcb-xi2pkk}.
 
 \sloppypar
Thanks to the combination of plenty of decay modes and advanced analysis techniques,
the LHCb collaboration achieved an impressive overall precision on the CKM angle $\gamma$
as shown in Fig.~\ref{fig:5}.
Including the new results presented here, the LHCb average~\cite{gamma} yields $\gamma = (67\pm4)^\circ$.
This is in excellent agreement with global CKM fits.
With the upcoming Run 3 data-taking
period, the combination of LHCb results will enter the high precision region where
discrepancies between direct measurement and indirect CKM prediction may be
observed. An ultimate precision at the sub-degree level will be achievable in the high luminosity LHC era. 
It remains thrilling to see whether new
phenomena beyond the established theory can be uncovered.
The anomaly observed in $B \to K\pi$ decays, strengthened by recent LHCb results,
might already point to
internal inconsistencies of the Standard Model.

\FloatBarrier

\section*{References}
\vspace{-0.5\baselineskip}


\begin{thebibliography}{99}

\bibitem{lhcb_b2dk_2body}LHCb collaboration, R. Aaij et al {\it et al}, \Journal{\JHEP}{04}{081}{2021}.

\bibitem{lhcb_b2dk_3body}LHCb collaboration, R. Aaij et al {\it et al}, \Journal{\JHEP}{02}{169}{2021}.

\bibitem{lhcb_b2kpi}LHCb collaboration, R. Aaij et al {\it et al}, \Journal{\PRL}{126}{091802}{2021}.



\bibitem{lhcb_bsdskpipi}LHCb collaboration, R. Aaij et al {\it et al}, \Journal{\JHEP}{03}{137}{2021}.

\bibitem{lhcb_dms}LHCb collaboration, R. Aaij et al {\it et al}, arXiv:2104.04421.

\bibitem{lhcb_phis}LHCb collaboration, R. Aaij et al {\it et al}, LHCB-PAPER-2020-042.

\bibitem{lhcb_b2hh}LHCb collaboration, R. Aaij et al {\it et al}, \Journal{\JHEP}{03}{075}{2021}.


\bibitem{lhcb-xi2pkk}LHCb collaboration, R. Aaij et al {\it et al}, arXiv:2104.15074.

\bibitem{gamma}LHCb collaboration, R. Aaij et al {\it et al}, LHCb-CONF-2020-003.






\end{thebibliography}
\end{document}